# Atomic adsorption on graphene with a single vacancy: systematic DFT study through the Periodic Table of Elements

Igor A. Pašti[1]*, Aleksandar Jovanović[1,2], Ana S. Dobrota[1], Slavko V. Mentus[1,3], Börje Johansson[4], Natalia V. Skorodumova[4,5]

[1]*University of Belgrade – Faculty of Physical Chemistry, Belgrade, Serbia*

[2]*CEST Kompetenzzentrum für elektrochemische Oberflächentechnologie GmbH, Wiener Neustadt, Austria*

[3]*Serbian Academy of Sciences and Arts, Belgrade, Serbia*

[4]*Department of Physics and Astronomy, Uppsala University, Uppsala, Sweden*

[5]*Department of Materials Science and Engineering, School of Industrial Engineering and Management, KTH - Royal Institute of Technology, Stockholm, Sweden*

---

* **corresponding author,** e-mail: igor@ffh.bg.ac.rs

**Vacancies in graphene present sites of altered chemical reactivity and open possibilities to tune graphene properties by defect engineering. The understanding of chemical reactivity of such defects is essential for successful implementation of carbon materials in advanced technologies. We report the results of a systematic DFT study of atomic adsorption on graphene with a single vacancy for the elements of rows 1 to 6 of the Periodic Table of Elements (PTE), excluding lanthanides. The calculations have been performed using PBE, long-range dispersion interaction-corrected PBE (PBE+D2 and PBE+D3) and non-local vdW-DF2 functional. We find that most elements strongly bind to the vacancy, except for the elements of groups 11 and 12, and noble gases, for which the contribution of dispersion interaction to bonding is most significant. The strength of the interaction with the vacancy correlates with the cohesive energy of the elements in their stable phases: the higher the cohesive energy is the stronger bonding to the vacancy can be expected. As most atoms can be trapped at the SV site we have calculated the potentials of dissolution and found that in most cases the metals adsorbed at the vacancy are more "noble" than they are in their corresponding stable phases.**

## 1. Introduction

Graphene has many intriguing electronic, thermal and mechanical properties that make it attractive for numerous applications.[1] In real life graphene typically has various types of defects and functional groups attached to it. Such defects reduce the performance of devices relying on the intrinsic properties of perfect graphene.[2] On the other hand, due to such defects the application of graphene in fields where pristine graphene is not suitable[2-4] becomes possible. A number of practical applications for such "imperfect" graphene are considered, such as, for example, spintronics and catalysis.

Various types of defects in graphene can be classified as intrinsic and extrinsic, while their dimensionality can also be different.[2] A single vacancy (SV) is the simplest defect in graphene, which is formed by removing one atom from the lattice. According to the literature around 7.5 eV is needed to form a single vacancy, the barrier for vacancy diffusion is estimated to be 1.3 eV while its mobility has been observed already at temperatures around 200°C.[5,6] When a carbon atom is removed it leaves unsaturated dangling bonds, which make the SV site very reactive.[3] Studies have shown that different atoms readily adsorb at the SV site.[7-9] In particular, Krasheninnikov *et al*.[8] have demonstrated strong binding of several 3d metals and Au to the SV site. Santos *et al*.[9] have performed a detailed theoretical analysis of the binding of the 3d metals to the SV site and addressed some issues related to the modeling of these impurities. We notice that several practical applications of graphene with imbedded impurity atoms have been suggested. For example, impurities can serve as catalytic sites for oxygen reduction reaction.[10,11]

There are many systematic studies of adatom adsorption on pristine graphene[12-14] including our recent analysis based on PBE, long-range dispersion interactions corrected PBE and non-local vdW-DF2 DFT calculations.[15] Although computational studies dedicated to the adsorption of impurities at the SV sites of graphene can also be found in literature, a general picture regarding the reactivity of SV is still lacking. Moreover, there is no systematic analysis of atomic adsorption at SV sites covering the whole or at least a large part of the Periodic Table of Elements (PTE). However, such studies are necessary in order to better understand the reactivity of such defect sites and specific materials chemistry and physics rendered by the introduction of impurities. Moreover, the availability of systematic databases covering different materials properties can be very useful in the context of materials informatics where such data can be used to design new materials.[16,17] For this reason, here we report the results of our systematic analysis of atomic adsorption at a vacancy in the graphene basal plane. We apply four computational schemes: PBE, long-range dispersion interactions corrected PBE and non-local vdW-DF2 DFT calculations, and analyze the trends in the formation of substitutional impurities in the graphene basal plane for the elements of the PTE up to atomic number 86, excluding lanthanides.

## 2. Computational details

We calculated the adsorption energies of all the elements of the PTE located in rows 1 to 6 (except lanthanides) at the SV site of graphene. Graphene sheet was modelled using a 4×4 cell (32 atoms). The repeated graphene sheets were separated from each other by 20 Å of vacuum. The effects of the cell size were tested and it was found that the 4×4 cell is sufficiently large to provide adequate results for atom adsorption at the SV site. This agrees with the result presented by Santos *et al*.[9]

The first-principle DFT calculations were performed using the Vienna *ab initio* simulation code (VASP).[18-20] The Generalized Gradient Approximation (GGA) in the parametrization by Perdew, Burk and Ernzerhof[21] combined with the projector augmented wave (PAW) method was used.[22] A cut-off energy of 600 eV and Gaussian smearing with a width of $\sigma$ = 0.025 eV for the occupation of the electronic levels were used. A Monkhorst-Pack $\Gamma$-centered 10×10×1 k-point mesh was used. Foreign atoms were placed at the SV site and during structural optimization the relaxation of all atoms in the simulation cell was unrestricted. The relaxation proceeded until the Hellmann-Feynman forces acting on all the atoms became smaller than $10^{-2}$ eV Å$^{-1}$.Spin-polarization was taken into account in all calculations.

To include dispersion interactions, which are not accounted for in PBE, we used different approaches. In the first step, we used DFT theory corrected for the long-range dispersion correction in the DFT+D2 and DFT+D3 formulations of Grimme.[23,24] Both approaches correct the total energy by a pairwise term ($E_{\mathrm{disp}}^{\mathrm{Dx}}$, $x$ = 2 or 3), which accounts for dispersion interaction and is added to the total energy of the system calculated using a selected DFT functional (in this case PBE):

$$E_{\mathrm{PBE+Dx}} = E_{\mathrm{DFT-PBE}} + E_{\mathrm{disp}}^{\mathrm{Dx}} \qquad (1)$$

$E_{\mathrm{disp}}^{\mathrm{Dx}}$ is obtained by the summation using the atom-specific parameters and relative distances over the entire simulation cell. For DFT+D2 the default set of parameters (as implemented in VASP) for the elements in rows 1-5 was used. For the elements of the 6$^{\mathrm{th}}$ row we used DFT+D2 parameters as described in Ref. ([15]).

In addition to the empirical correction for dispersion interactions, we applied the vdW-DF2 non-local functional developed by Langreth's and Lundqvist's groups.[25] The method is implemented in VASP in a way allowing for the inclusion of the non-local contribution into the correlation energy during the self-consistency cycle.[26] The accuracy of non-local functionals increases when higher number of valence electrons is taken into account so we used PAW potentials with semi-core states where possible.

The energy of the atom adsorption at the SV site is quantified as atomic adsorption energy at the SV site and is calculated as:

$$E_{\mathrm{ads}}^{\mathrm{PBE}} = E_0^{\mathrm{PBE}}[\mathrm{A@SV\text{-}G}] - E_0^{\mathrm{PBE}}[\mathrm{SV\text{-}G}] - E_0^{\mathrm{PBE}}[\mathrm{A}] \quad (2)$$

$$E_{\mathrm{ads}}^{\mathrm{PBE+D}} = E_0^{\mathrm{PBE+D}}[\mathrm{A@SV\text{-}G}] - E_0^{\mathrm{PBE+D}}[\mathrm{SV\text{-}G}] - E_0^{\mathrm{PBE}}[\mathrm{A}] \quad (3)$$

$$E_{\mathrm{ads}}^{\mathrm{vdW\text{-}DF2}} = E_0^{\mathrm{vdW\text{-}DF2}}[\mathrm{A@SV\text{-}G}] - E_0^{\mathrm{vdW\text{-}DF2}}[\mathrm{SV\text{-}G}] - E_0^{\mathrm{vdW\text{-}DF2}}[\mathrm{A}] \quad (4)$$

where $E_0$ are the ground state energies of the adatom adsorbed at the SV site of graphene [A@SV-G], graphene with SV [SV-G] and isolated atom [A], calculated with the method specified in superscript. $E_{\mathrm{ads}}$ is negative when adsorption is exothermic.

## 3. Results and discussion

### *3.1. Graphene with monovacancy*

For each applied scheme the graphene lattice was fully optimized. The obtained C–C bond lengths are provided in Table 1. We further investigated the energetics of vacancy formation, using vacancy formation energies ($E_{\mathrm{vf}}$) defined as:

$$E_0^{\mathrm{method}} = E_0^{\mathrm{method}}[\mathrm{SV - G}] - \frac{n-1}{n} E_0^{\mathrm{method}}[\mathrm{G}] \quad (5)$$

where $E_0$[G] stands for the total energy of pristine graphene sheet while $n$ is the number of carbon atoms in the simulation cell ($n$ = 32). The calculated values are reported in Table 1. Good agreement with previous literature reports is obtained as the reported vacancy formation energies are in the range from 7.2 to 8.58 eV.[6,27] It is recognized that the removal of a carbon atom from the graphene lattice gives rise to magnetic behavior. Ma *et al*.[28] reported the magnetic moment of 1.04 $\mu_B$ using a 128-atom graphene supercell. According to the overview by Valencia and Caldas[29] this is the lowest among the reported values while the highest one is 2 $\mu_B$. As it has been shown by Rodrigo *et al*.[30] an accurate theoretical prediction of the magnetic moment of SV-graphene requires a rigorous treatment of the system as the value of the magnetic moment is very sensitive to the size of the simulation cell and the density of the $k$-point mesh.[30] A highly converged (with respect to the $k$-point mesh) value of magnetic moment obtained for the 6×6 cell is around 1.6 $\mu_B$, and it increases with the lateral size of the supercell. In particular, for the 30×30 cell it is close to 1.75 $\mu_B$. The density of the k-point mesh applied in our study is expected to provide an adequate estimate of the magnetic properties of SV-graphene, while the calculated energies, which are the focus of the present work, are highly converged. The calculated densities of states (DOS) of SV-graphene (Fig. 1, Fig. S1) show the semi-localized π states of vacancy in the vicinity of the Fermi level, in agreement with previous reports.[30] Based on the comparison of the obtained results and the previous literature reports we conclude that our model is adequate for the analysis of atomic adsorption at the SV site.

**Table 1**. Calculated C–C bond lengths, vacancy formation energies and ground state magnetic moments of SV-graphene sheet in 4×4 cell using different computational schemes.

| Computational scheme | C–C bond length[a] / Å | vacancy formation energy / eV | magnetization / $\mu_B$ |
|---|---|---|---|
| **PBE** | 1.425 | 7.768 | 1.36 |
| **PBE+D2** | 1.425 | 7.815 | 1.35 |
| **PBE+D3** | 1.425 | 7.778 | 1.35 |
| **vdW-DF2** | 1.430 | 7.389 | 1.35 |

[a]ref. ([15])

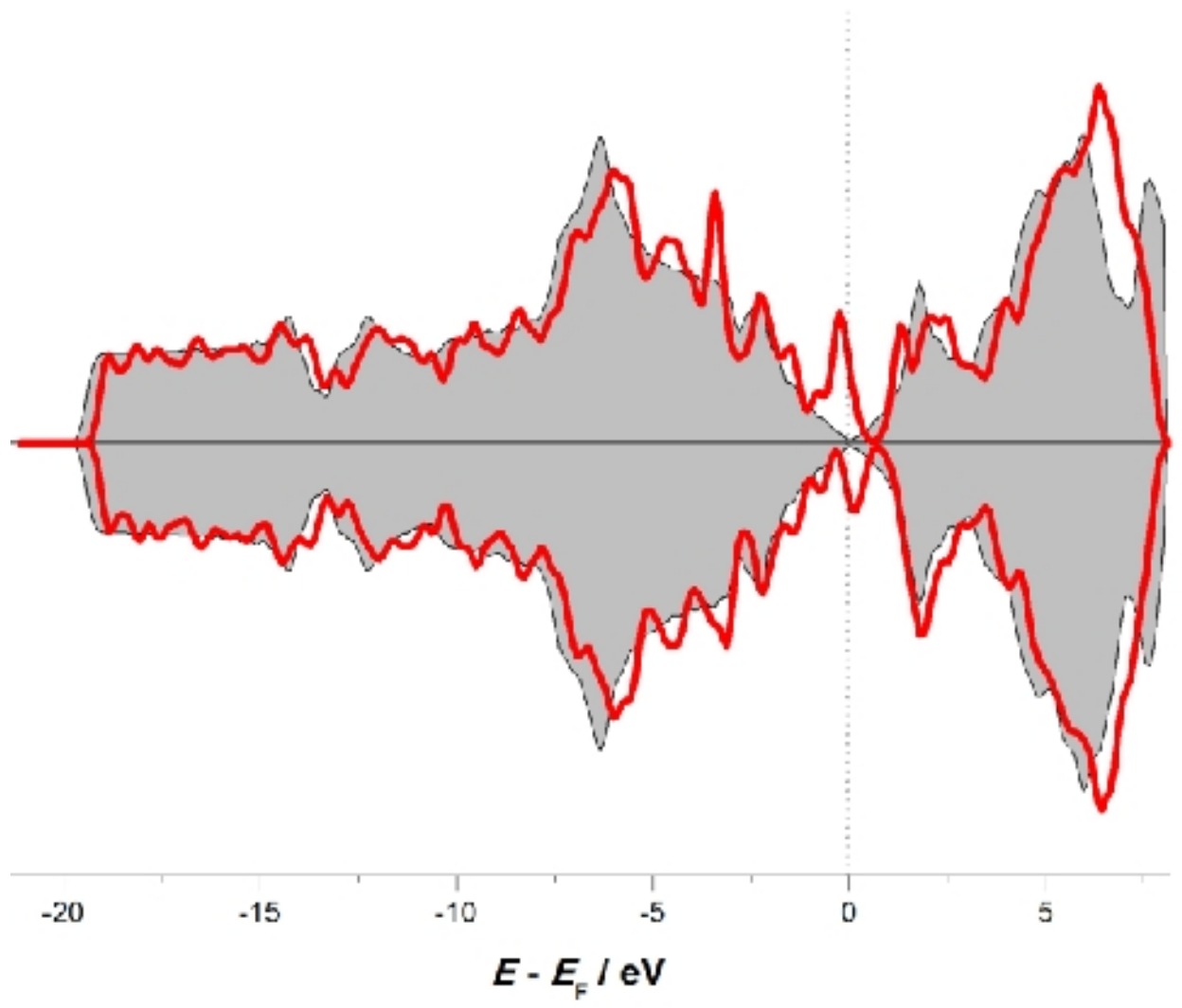

**Figure 1**. DOS plots for SV-graphene (thick line) superimposed on DOS of pristine graphene (shaded area) obtained using PBE approach.

*3.2. Atomic adsorption at single vacancy sites*

To study atomic adsorption we used different starting positions for the impurity atom with respect to the SV site of graphene. For the final relaxed structures we find that a large portion of the elements prefers the configuration with the $C_{3v}$ symmetry, where the adatom is situated in the middle of the vacancy forming three equal impurity-C bonds. Due to differences in atomic sizes, most of the atoms are located above the carbon plane, but some, like B or N, practically substitute the C atom. Such a doped graphene sheet remains perfectly flat that also agrees with previous reports.[3] On the other hand, some elements do prefer bonding in lower symmetry configurations. Among them are Zn (as also shown previously by Santos *et al.*[9]), Cd, Hg, H, O and halogen elements. For Cl we find that solutions with different magnetic moments can be stabilized depending on the number of Cl-C bonds formed, while the ground state is the configuration where only one Cl-C bond is formed. Some representative examples of different bonding configuration at the SV site are shown in Fig. S2 (Supplementary Information), while the full list of bonding distances is presented in Table S1.

The calculated energies of atomic adsorption at the SV site are presented in Figs. 2-5 and all the data are assembled in Table S2 (Supplementary Information). The calculated magnetic moments of the A@SV-graphene systems are shown and discussed in Supplementary Information, Table S3.

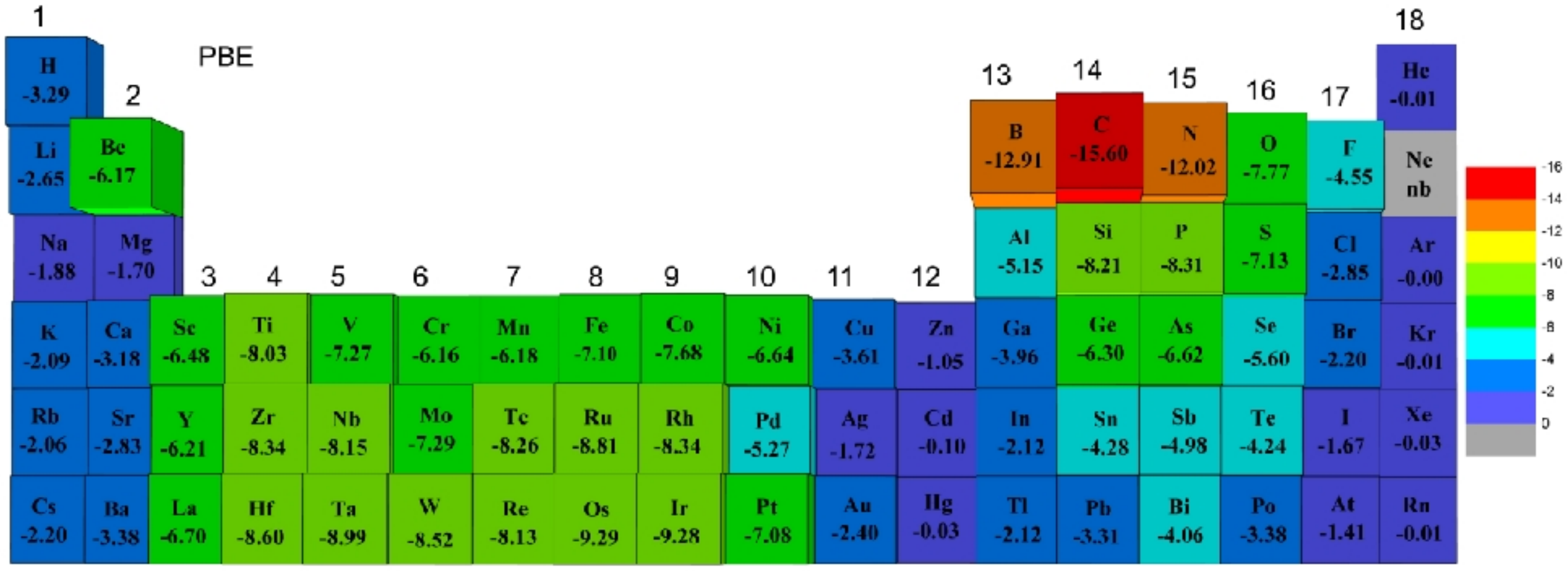

**Figure 2.** Calculated adsorption energies (in eV) for atomic adsorption at SV using PBE calculations. Group number is indicated.

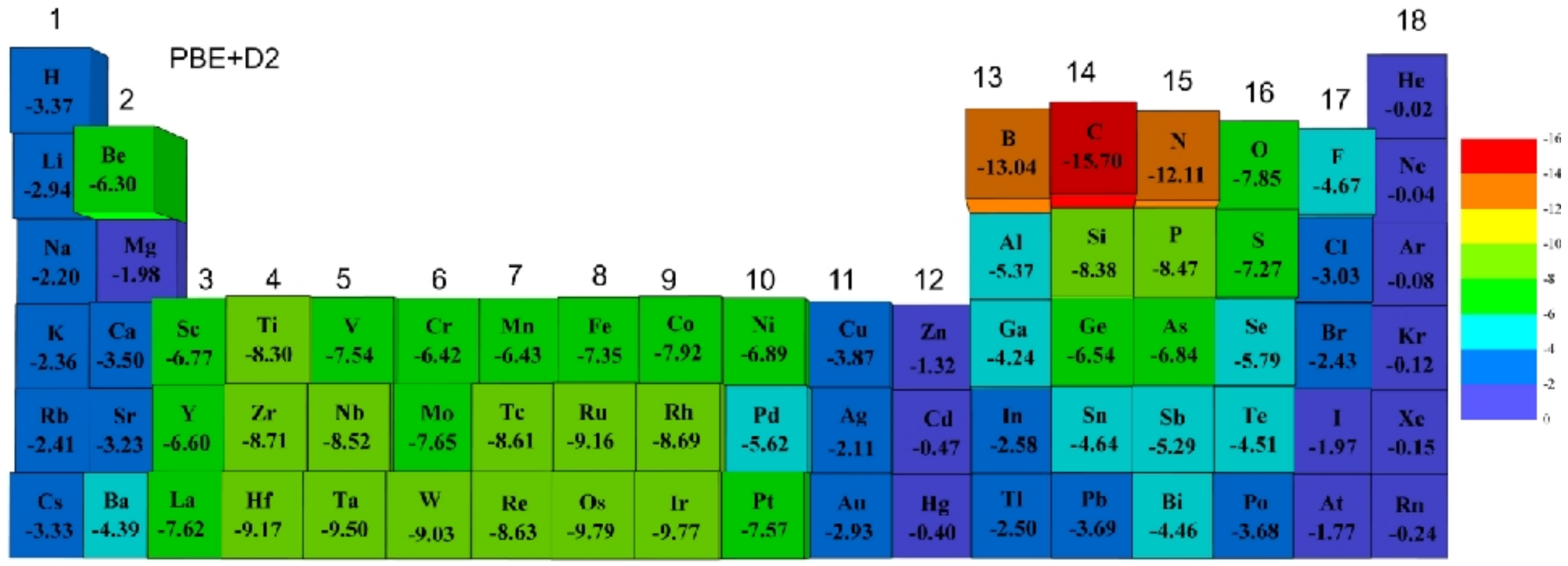

**Figure 3.** Calculated adsorption energies (in eV) for atomic adsorption at SV using PBE+D2 calculations. Group number is indicated.

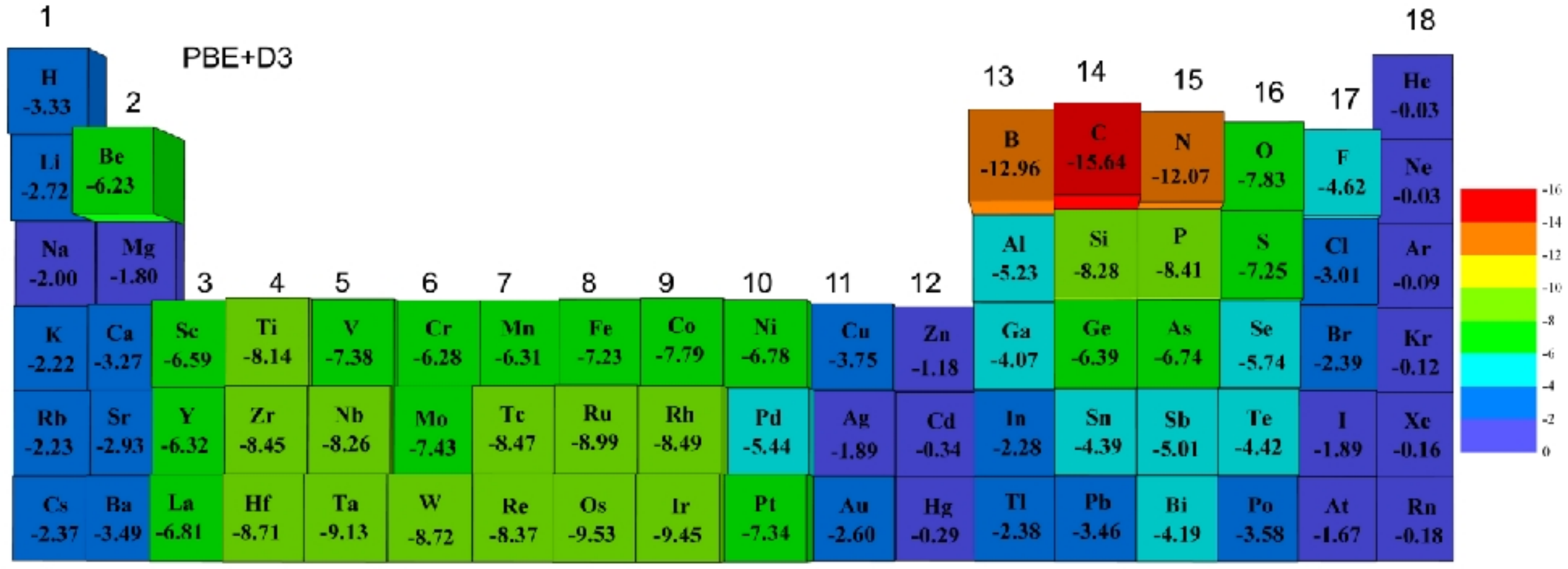

**Figure 4.** Calculated adsorption energies (in eV) for atomic adsorption at SV using PBE+D3 calculations. Group number is indicated.

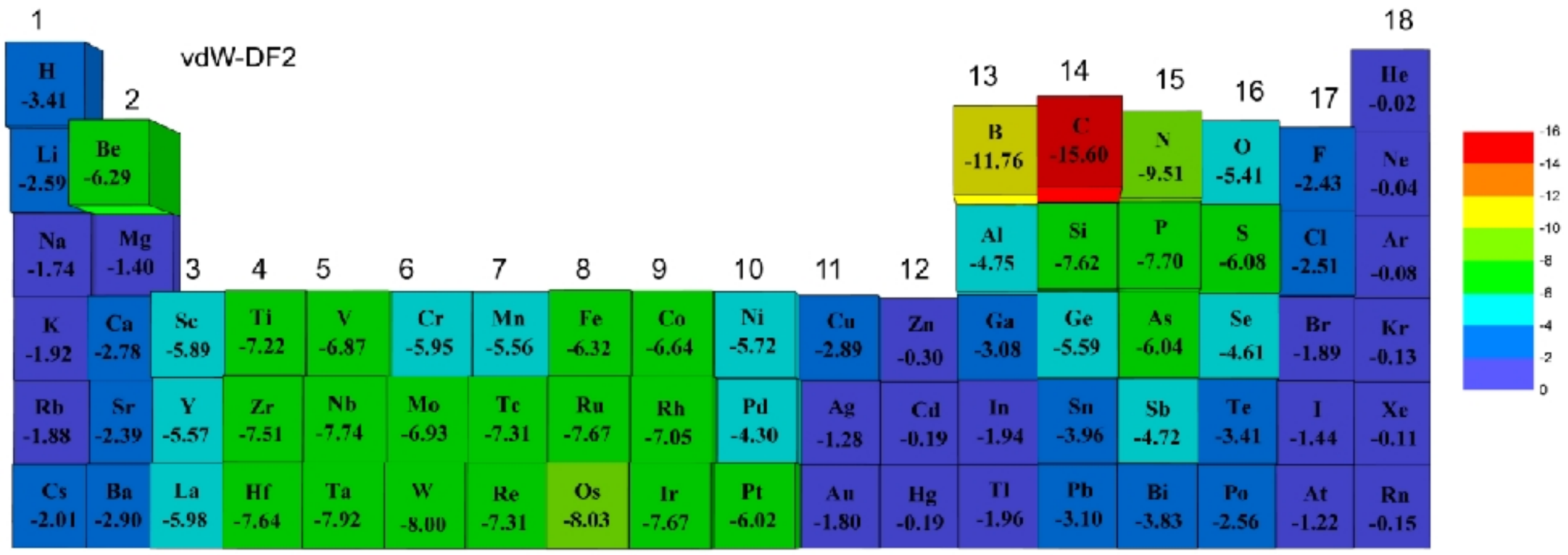

**Figure 5.** Calculated adsorption energies (in eV) for atomic adsorption at SV using vdW-DF2 calculations. Group number is indicated.

As can be seen, the four applied computational schemes give similar overall trends in the adsorption energies. In fact, there is a very good correlation between the PBE results and the results of the other three computational methods (Fig. 6). Interestingly, the correlation is much better than that obtained for atomic adsorption on pristine graphene.[15] while the order of the interaction strength is the same: PBE+D2 > PBE+D3 > PBE > vdW-DF2.

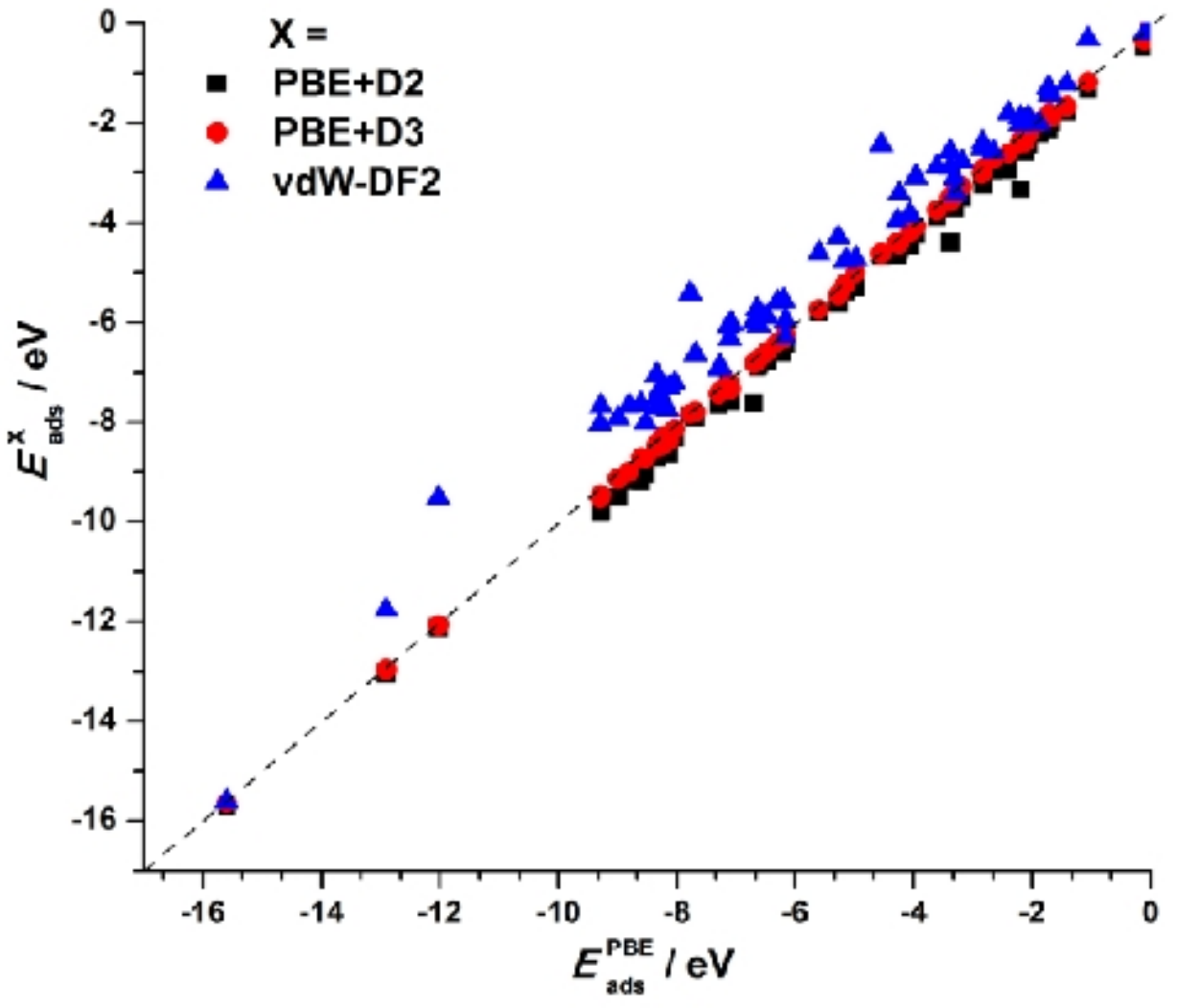

**Figure 6.** Correlation between PBE results and PBE+D2 (■), PBE+D3 (●) and vdW-DF2 (▲) results for the energy of adsorption of adatom at SV sites of graphene.

This suggests that in the case of single vacancies a strong chemical interaction dominates bonding and that the contribution of dispersion interaction is relatively small. As expected, the highest adsorption energy was found for C followed by B and N, which are truly embedded into the carbon lattice. The *d*-elements interact rather strongly with the SV site with energies typically between −8 eV and −6 eV. We note that the obtained PBE results along the series of the 3d elements are in agreement with those reported in Ref. ([9]). The weakest interactions are seen for the elements with closed shells, corresponding to configurations $ns^2(n-1)d^{10}$ and $ns^2np^6$ (noble gases). This is understandable as these elements tend to keep their shells closed. Regarding the trends along the rows of the PTE, we observe shallow minima for the elements with half-filled d-band and pronounced minima for the elements having $ns^2(n-1)d^{10}$ configurations. In contrast to the adsorption on pristine graphene, the elements with the $np^3$ configuration (group 15 of the PTE) strongly bind to the vacancy using the dangling bonds at the SV site to maximally fill their valence bands. This can be explained by the $sp^3$ hybridization of these elements, which results in an orbital whose symmetry corresponds to that of the dangling bonds at the vacancy site, providing a maximum overlap and strong chemical interaction. Considering trends along the groups of the PTE, for s- and p-elements the strength of the interaction typically decreases when going down the groups. For the case of the elements with partially filled d-shell and noble gases the adsorption at the SV site becomes stronger going down the group.

To get a better insight into the role of dispersion interactions in binding of different elements we calculated the percentage of dispersion contribution to $E_{ads}$ in the PBE+D2 scheme with respect to that in PBE (Fig. 7). We did the same also for the adsorption on pristine graphene using our previously reported[15] values (Fig. 7). We see that the contribution of dispersion to the binding energies is maximized for weakly interacting elements (groups 12 and 18). In the case of pristine graphene the same trend is seen for groups 6, 7 and 15, whose elements physisorb on pristine graphene. The same conclusions regarding the role of dispersion interactions hold in the case of PBE+D3 (not presented here), although the contribution of dispersion is typically lower compared to that of PBE+D2. For adsorption at SV we also see that along the groups of the PTE the contribution of dispersion typically increases with the atomic number. Considering small relative contribution of dispersion obtained in PBE+D2 and PBE+D3 and a good scaling between them and the PBE results (Fig. 6), we conclude that for an adequate modelling of atomic adsorption, focusing on the energetics of the process, PBE is sufficient in most cases.

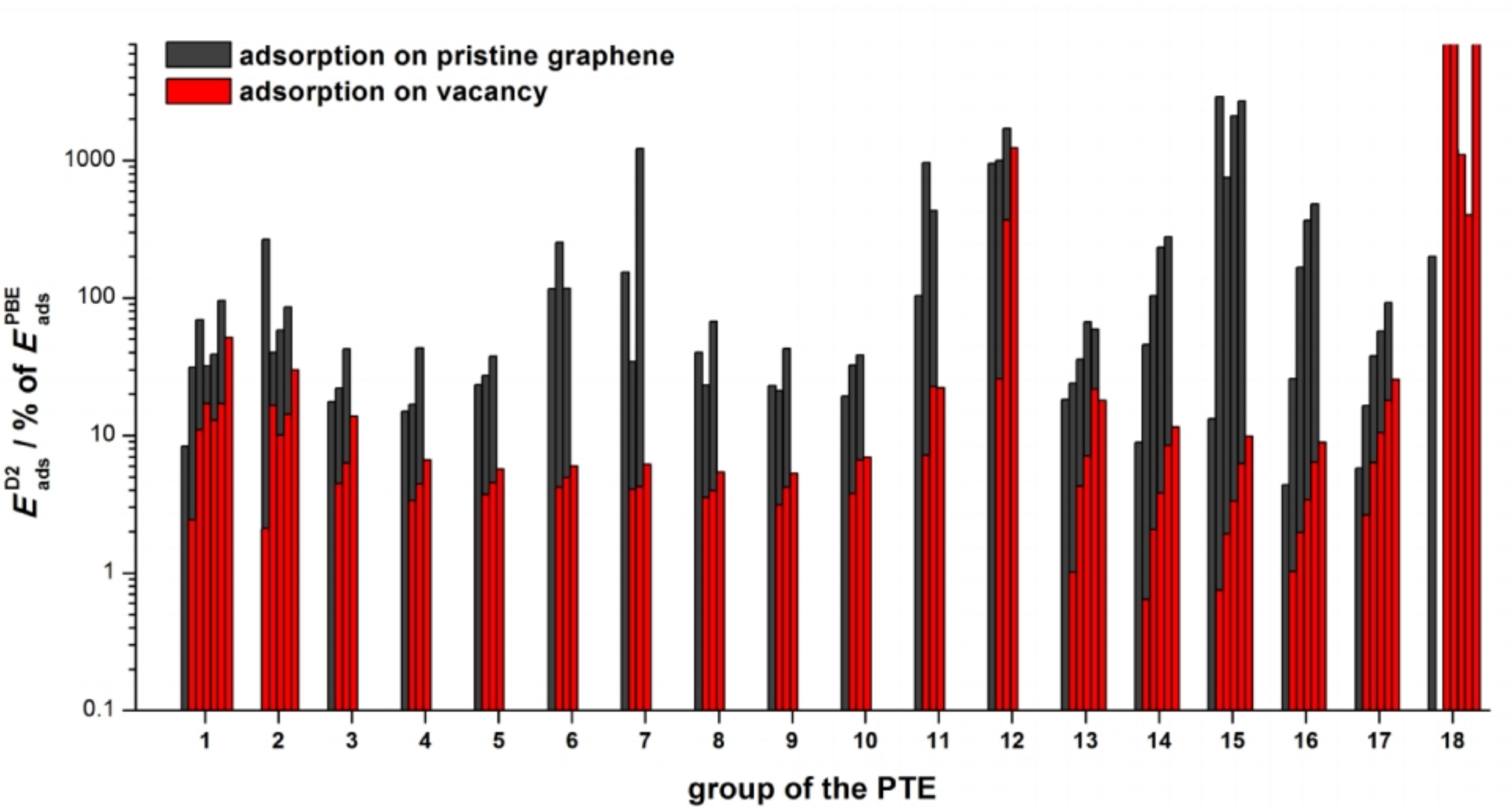

**Figure 7**. Relative contribution of dispersion interactions to the adsorption energies within the PBE+D2 scheme with respect to PBE. For each group of the PTE elements are sorted so that the atomic number increases from left to right.

Further, we correlate calculated $E_{ads}$ with the experimental values of cohesive energies in the stable phases of pure elements.[31] Such a correlation is shown in Fig. 8 for the case of vdW-DF2 and it holds for other applied schemes as well. Having analyzed the calculated set of data, we see that $E_{ads}$ at SV correlates well with the cohesive energies providing a simple rule of thumb: the stronger the bonding of the element in its stable phase is the stronger its interaction with the SV site will be. Previously, similar trends in the behaviour of cohesive and adsorption energies at SV were observed by Chen *et al*.[11] for a limited number of elements. We note that we observe no correlation between the adsorption energies on pristine graphene and on SV-graphene (Fig. S4, Supplementary Information). Accordingly, we neither find any correlation between the cohesive energies of the elements and their adsorption energies on pristine graphene. Such a difference between atomic adsorption on pristine graphene and SV-graphene originates from different bonding occurring in these two cases. In the latter case true chemical interactions take place, while in the case of pristine graphene the interaction is achieved through the π electron system which in many cases does not provide real chemical bonding. Fig.8 shows that the elements can be sorted into two groups. In particular, the atoms with partially filled d-shells can be grouped together as they show relatively small sensitivity to the increase of the cohesive energy of pure metallic phase. In contrast, the elements with filled d-shells (or the ones without d-electrons) demonstrate a strong correlation between the cohesive energy and $E_{ads,}$ when going down the groups in the PTE.

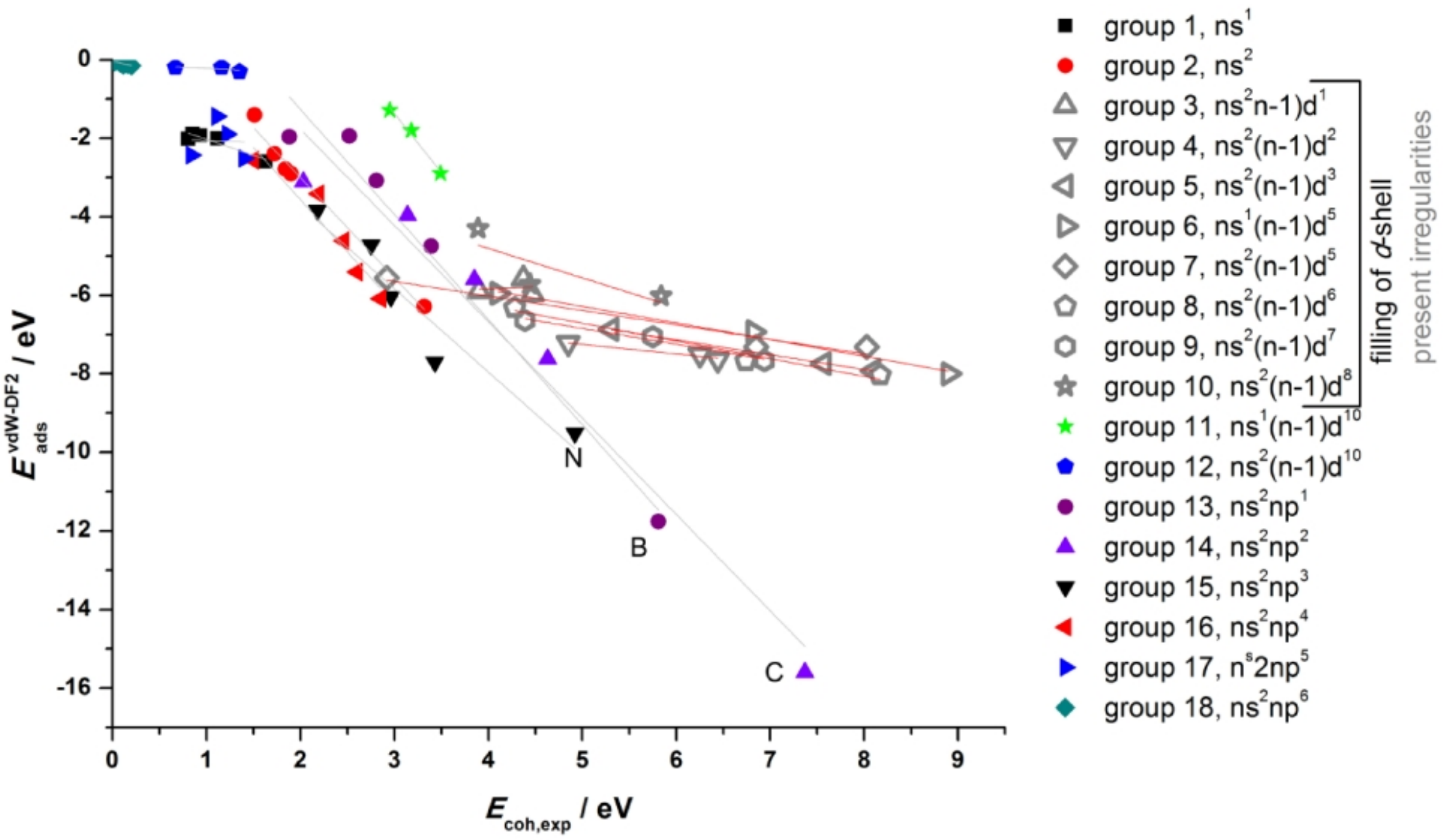

**Figure 8**. Correlation between the cohesive energy and $E_{\mathrm{ads}}^{\mathrm{vdW\text{-}DF2}}$

Comparison of adsorption and cohesive energies can also help us to understand whether atoms prefer to adsorb at SV or rather bind together and form clusters over graphene. First, all the elements interact significantly weaker with pristine graphene than with the SV site, so a separate atom would prefer to bind to the SV site rather than to sites of pristine graphene basal plane. As the diffusion barrier for adatoms on graphene is relatively small[14] atoms can migrate and either be trapped at SV or agglomerate to form clusters. To estimate the tendency of the latter process we compare the experimental cohesive energies to the calculated adsorption energies at SV in Fig. 9. We see that all the four applied computational schemes predict that the majority of the elements will be trapped at the vacancy, while some of them could prefer to form bonds with other atoms of their kind. The latter group consists of the 5d elements with exceptionally high cohesive energy (W being the most prominent example) and the elements of groups 11 (Cu, Ag and Au) and 12 (Zn, Cd, Hg), which rather weakly interact with SV.

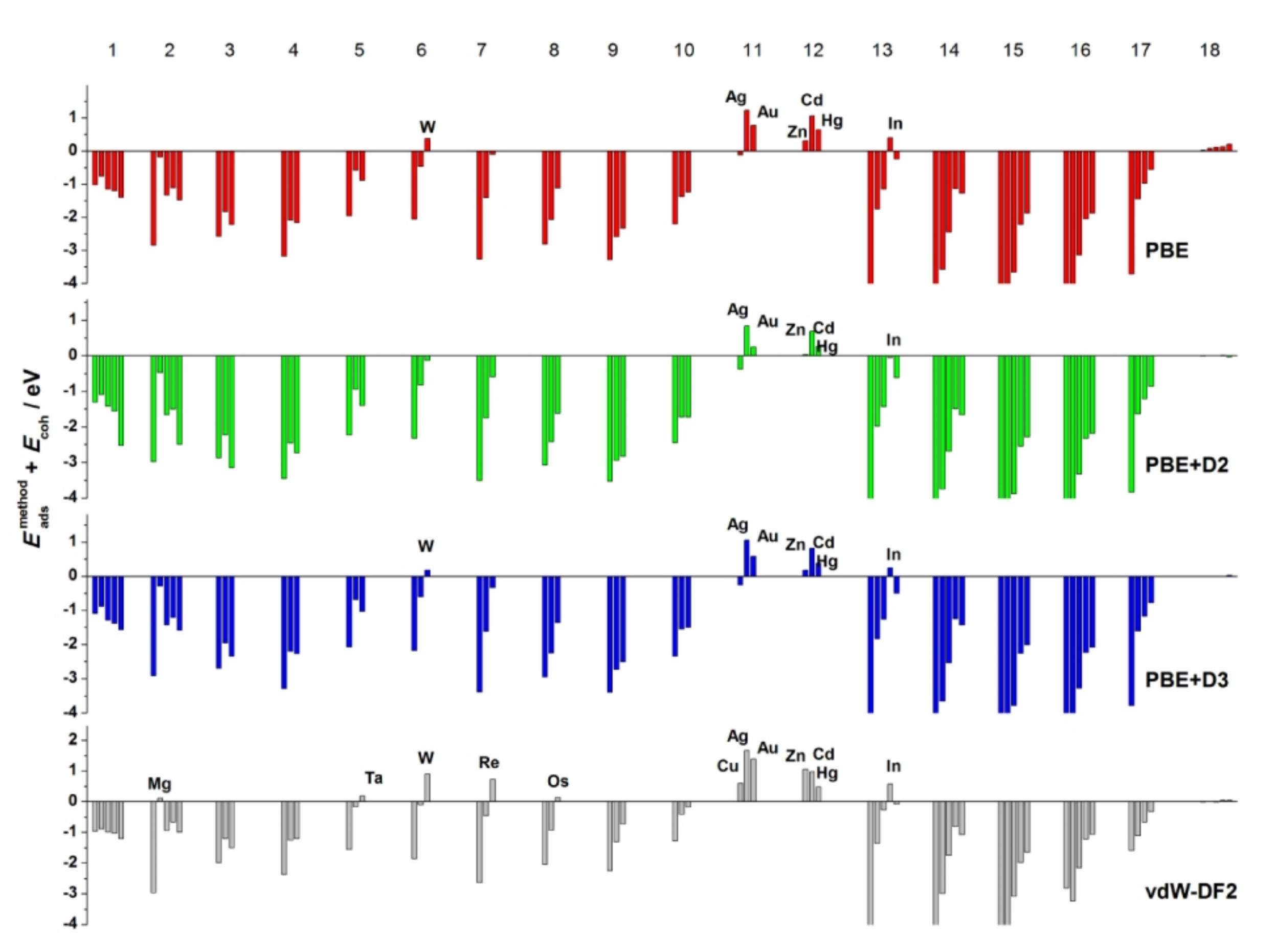

**Figure 9**. $E_{ads}$ at SV site compared to the cohesive energies of elements in their pure phases. Negative values indicate that given elements are more stable at SV than in their pure phases. The results are provided for all four applied computational scheme. Top row of numbers gives the group of the PTE.

The observed stabilization of metal atoms at SV as compared to their pure phases (Fig. 9) can be connected to the tendency of these atoms to dissolve from SV-graphene. This is of great importance when one considers possible applications of M@SV-graphene systems as catalysts in wet processes, where dissolution and corrosion can take place. By adapting the approach described in Ref. ([4]) we calculated the standard electrode potentials of the following process:

$$M^{z+} + \text{SV-graphene} + ze^{-} \rightarrow \text{M@SV-graphene} \quad (6)$$

for all the investigated metals. This actually gives the dissolution potential for considered impurities in graphene lattice. The data are provided in Table S4, Supplementary Information. We considered that our SV-graphene and M@SV-graphene were in their standard states as being solid phases. Based on these results we considered the dissolution of M from the SV site of graphene basal plane under different pH conditions and found that, when adsorbed at SV, certain elements should not dissolve according to Eq. (6), followed by the evolution of $H_2$, irrespective of pH (standard potential above 0 V *vs*. SHE, Fig. 10). It can be concluded that these impurities are less prone to dissolution than their metallic phases (in other words, they are more "noble" than the corresponding pure metals), except for the metals listed in Fig. 10. This is the case for the majority of catalytically interesting metals located in the d-block of the PTE. Among them are, for example, Fe, Co and Ni, which are very prone to corrosion in acidic solutions. In contrast, some elements should dissolve from the SV site at every pH and such M@SV-graphene systems cannot be considered stable in aqueous media (standard potential below −0.826 V). For the intermediate cases an adequate pH range can be found where the material can be used without dissolution of the metal. Considering electrochemical applications there is an additional factor, namely, the electrode potential. In that case for each metal adsorbed at SV and each pH one can find a potential window where M@SV-graphene is stable. Also for such a scenario, there are metals, which do not dissolve from the SV site within a (theoretical) potential window available in aqueous solutions (electrode potential above 1.23 V *vs*. SHE: Ru, Co, Rh, Pd, Pt, Au). The presented results can be used for a quick pre-screening of possible candidates for a given (electro)catalytic reaction under given conditions. For example, Kaukonen *et al*. among other systems proposed Sn adsorbed at SV as a possible oxygen reduction reaction (ORR) catalyst.[10] However, the dissolution potential of Sn@SV-graphene is around 0.5 V *vs*. SHE and it is not to be stable under acidic conditions at the potentials where ORR takes place. It is, however, stable under highly alkaline conditions at the potentials of ORR. If some other species are present in the solution, which can enhance dissolution, like chloride or some other complexing agents, this can also be taken into account through the use of Nernst equation. However, in that case it is also possible to evaluate dissolution potentials using calculated adsorption energies at the SV site.

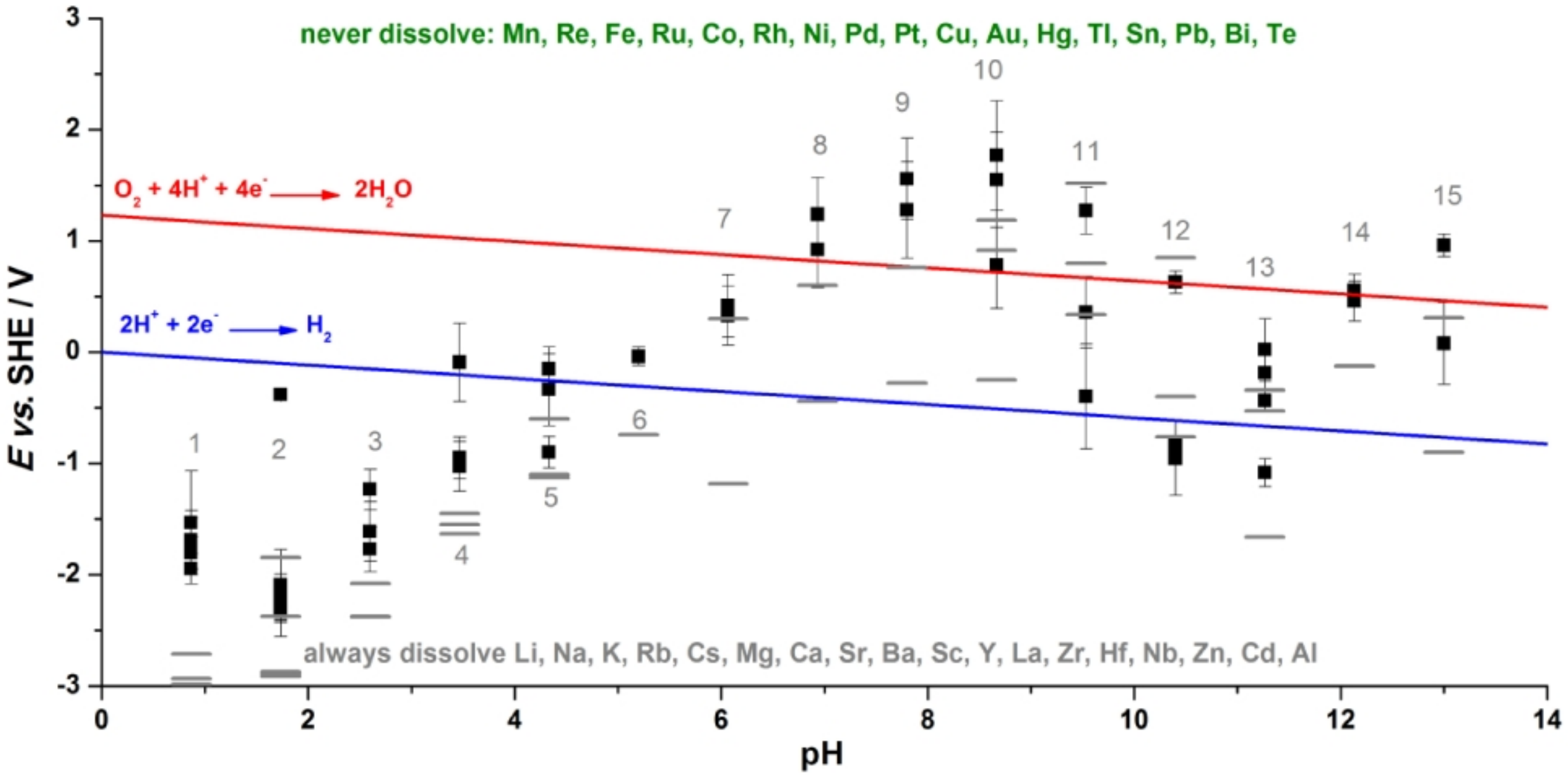

**Figure 10**. Average electrode potentials for the dissolution of metal adsorbed at the SV site of graphene. Numbers next to symbols denote the group of the PTE. Data points give the average electrode potential obtained using four computational schemes while the error bars indicate their variation among the used methods. Note that the data points for the electrode potentials of the considered metals are not linked to pH scale. Horizontal lines give electrode potentials for the pure metallic phases of given elements.

## 4. Conclusions

We have analyzed the adsorption of all the elements of the PTE up to atomic number 86 (excluding lanthanides) at the SV of graphene sheet. We find a strong interaction of the majority of the elements with the SV site. Relatively weak interactions are seen only for the elements of groups 11 (Cu, Ag and Au), 12 (Zn, Cd and Hg) and 18 (noble gases). We find a link between the cohesive energies of the elements in their pure phases and their adsorption energies at the SV site. Metal atom impurities adsorbed at SV sites are typically less prone to dissolution than their corresponding pure metallic phases. The provided results offer a comprehensive view on the reactivity of the SV site and stability of adsorbed impurities, which can be useful for designing advanced graphene-based materials for various applications.

## Conflicts of interest

There are no conflicts to declare.

## Acknowledgements

I.A.P., A.S.D. and S.V.M. acknowledge the support provided by the Serbian Ministry of Education, Science and Technological Development through the project III45014. N.V.S. acknowledges the support provided by Swedish Research Council through the project No. 2014-5993. The computations were performed on resources provided by the Swedish National Infrastructure for Computing (SNIC) at National Supercomputer Centre (NSC) at Linköping University. Financial support provided through the NATO Project EAP.SFPP 984925 - "DURAPEM - Novel Materials for Durable Proton Exchange Membrane Fuel Cells" is also acknowledged. We also acknowledge the support from Carl Tryggers Foundation for Scientific Research.

# SUPPLEMENTARY INFORMATION

## 1. Electronic structure of SV-graphene system

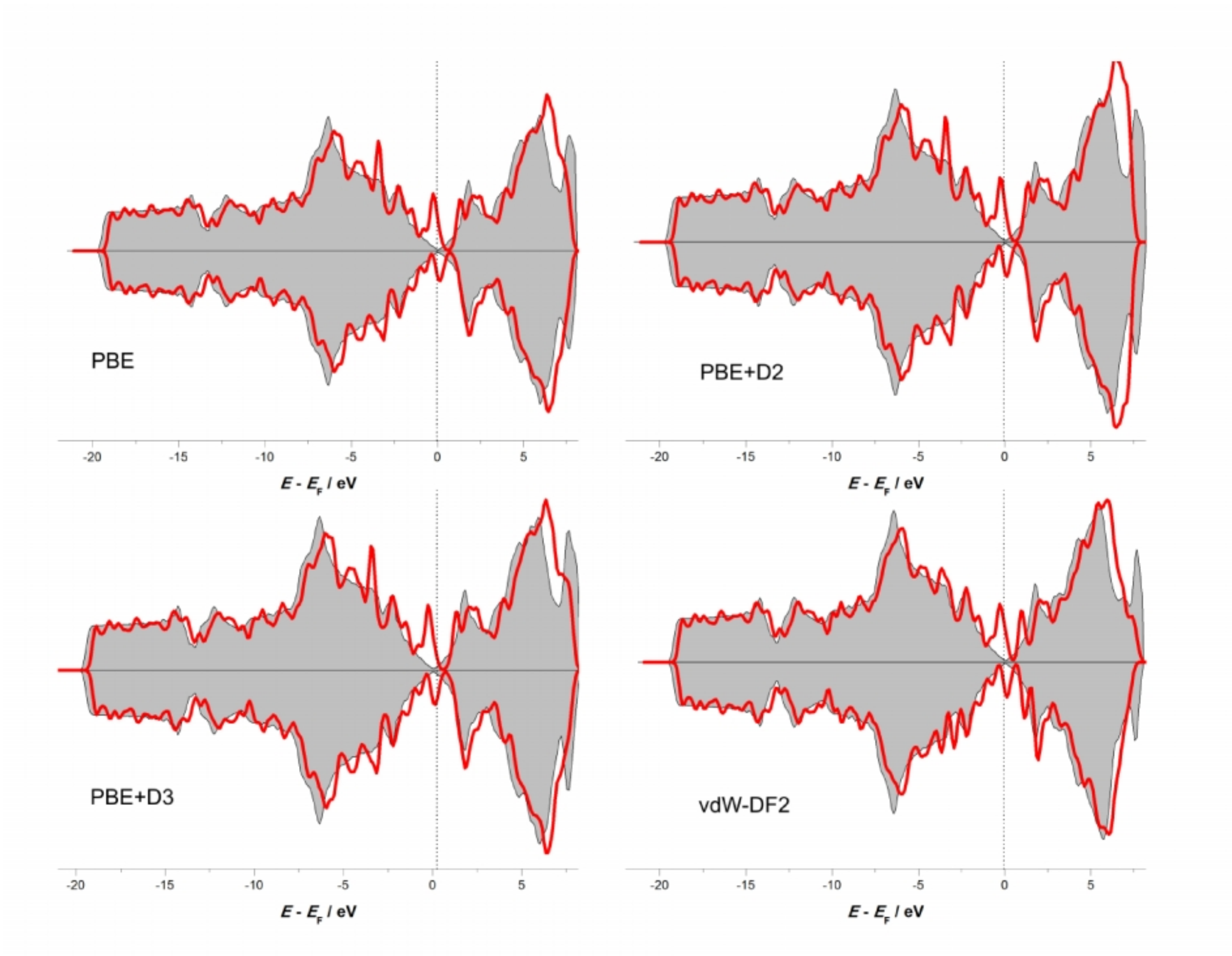

**Figure S1**. DOS plots for SV-graphene (thick line) superimposed on DOS of pristine graphene (shaded area) using designated computational schemes.

## 2. Bonding of elements from different parts of the PTE at the SV site of graphene

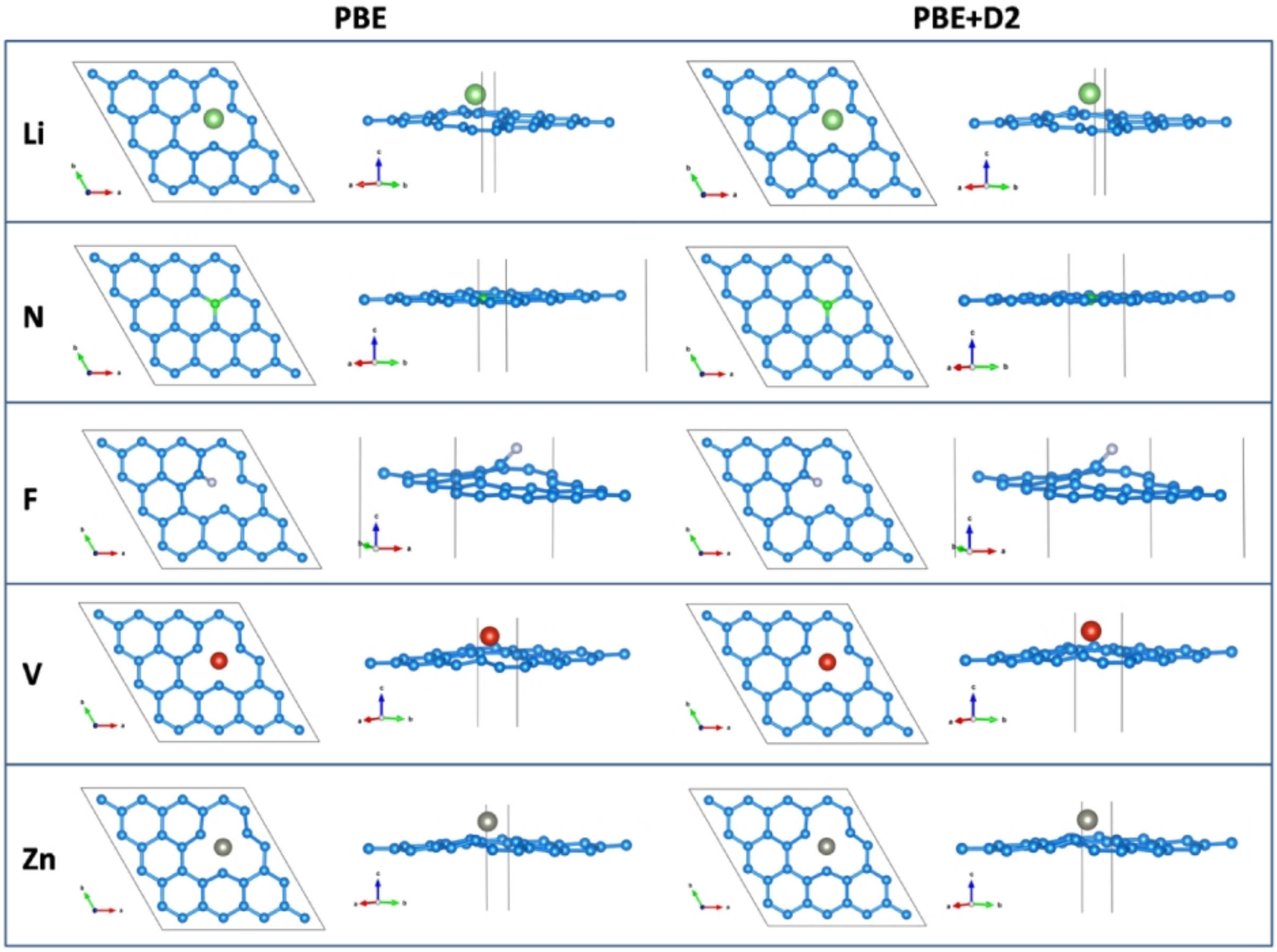

**Figure S2**. Some examples of the ground state geometries obtained using PBE and PBE+D2. Carbon atoms are blue, while designated impurity atoms are given in different colors.

**Table S1.** Calculated distances of atoms adsorbed at SV−graphene from the graphene plane (in Å)

| | | | | | | | | | | | | | | | | | |
|---|---|---|---|---|---|---|---|---|---|---|---|---|---|---|---|---|---|
| **H**<br>0.023<br>0.023<br>0.026<br>0.284 | | | | | **element**<br>PBE<br>PBE+D2<br>PBE+D3<br>vdW−DF2 | | | | | | | | | | | | **He**<br>3.274<br>3.038<br>3.037<br>2.988 |
| **Li**<br>1.687<br>1.730<br>1.690<br>1.583 | **Be**<br>0.842<br>0.856<br>0.855<br>0.798 | | | | | | | | | | | **B**<br>0.043<br>0.047<br>0.033<br>0.028 | **C**<br>0.000<br>0.000<br>0.000<br>0.000 | **N**<br>0.010<br>0.009<br>0.010<br>0.006 | **O**<br>0.315<br>0.316<br>0.316<br>0.327 | **F**<br>1.654<br>1.655<br>1.653<br>1.542 | **Ne**<br>2.877<br>2.705<br>2.763<br>2.782 |
| **Na**<br>2.142<br>2.112<br>2.169<br>2.160 | **Mg**<br>1.754<br>1.741<br>1.757<br>1.797 | | | | | | | | | | | **Al**<br>1.388<br>1.391<br>1.391<br>1.390 | **Si**<br>1.167<br>1.169<br>1.186<br>1.186 | **P**<br>1.153<br>1.164<br>1.164<br>1.191 | **S**<br>1.063<br>1.073<br>1.073<br>1.073 | **Cl**<br>2.136<br>2.139<br>2.135<br>2.170 | **Ar**<br>3.359<br>3.145<br>3.192<br>3.191 |
| **K**<br>2.491<br>2.436<br>2.528<br>2.503 | **Ca**<br>1.974<br>1.952<br>1.986<br>2.063 | **Sc**<br>1.735<br>1.742<br>1.744<br>1.787 | **Ti**<br>1.474<br>1.481<br>1.484<br>1.519 | **V**<br>1.399<br>1.418<br>1.419<br>1.465 | **Cr**<br>1.359<br>1.382<br>1.375<br>1.427 | **Mn**<br>1.293<br>1.297<br>1.297<br>1.327 | **Fe**<br>1.222<br>1.238<br>1.222<br>1.254 | **Co**<br>1.179<br>1.189<br>1.182<br>1.226 | **Ni**<br>1.192<br>1.195<br>1.195<br>1.236 | **Cu**<br>1.289<br>1.303<br>1.293<br>1.388 | **Zn**<br>1.445<br>1.456<br>1.439<br>1.606 | **Ga**<br>1.372<br>1.379<br>1.375<br>1.48 | **Ge**<br>1.427<br>1.436<br>1.434<br>1.542 | **As**<br>1.458<br>1.458<br>1.464<br>1.524 | **Se**<br>1.345<br>1.355<br>1.355<br>1.381 | **Br**<br>2.320<br>2.321<br>2.331<br>2.361 | **Kr**<br>3.710<br>3.304<br>3.362<br>3.380 |
| **Rb**<br>2.641<br>2.56<br>2.679<br>2.661 | **Sr**<br>2.135<br>2.107<br>2.151<br>2.252 | **Y**<br>1.939<br>1.940<br>1.950<br>1.997 | **Zr**<br>1.704<br>1.719<br>1.718<br>1.773 | **Nb**<br>1.62<br>1.634<br>1.63<br>1.68 | **Mo**<br>1.578<br>1.592<br>1.587<br>1.637 | **Tc**<br>1.514<br>1.521<br>1.510<br>1.571 | **Ru**<br>1.464<br>1.476<br>1.476<br>1.548 | **Rh**<br>1.460<br>1.439<br>1.439<br>1.520 | **Pd**<br>1.447<br>1.459<br>1.454<br>1.535 | **Ag**<br>1.746<br>1.763<br>1.763<br>2.033 | **Cd**<br>2.432<br>2.353<br>2.609<br>3.314 | **In**<br>2.107<br>2.103<br>2.135<br>2.268 | **Sn**<br>1.931<br>1.939<br>1.936<br>1.996 | **Sb**<br>1.819<br>1.856<br>1.853<br>1.863 | **Te**<br>1.694<br>1.696<br>1.694<br>1.731 | **I**<br>2.540<br>2.528<br>2.544<br>2.627 | **Xe**<br>3.657<br>3.170<br>3.347<br>3.178 |
| **Cs**<br>2.788<br>2.532<br>2.808<br>2.815 | **Ba**<br>2.269<br>2.203<br>2.278<br>2.373 | **La**<br>2.029<br>2.029<br>2.038<br>2.115 | **Hf**<br>1.665<br>1.669<br>1.665<br>1.721 | **Ta**<br>1.588<br>1.601<br>1.595<br>1.655 | **W**<br>1.584<br>1.587<br>1.585<br>1.617 | **Re**<br>1.540<br>1.556<br>1.546<br>1.585 | **Os**<br>1.512<br>1.517<br>1.517<br>1.554 | **Ir**<br>1.486<br>1.488<br>1.486<br>1.528 | **Pt**<br>1.467<br>1.482<br>1.482<br>1.526 | **Au**<br>1.637<br>1.659<br>1.648<br>1.817 | **Hg**<br>3.483<br>2.855<br>2.998<br>3.315 | **Tl**<br>2.436<br>2.435<br>2.552<br>2.501 | **Pb**<br>2.128<br>2.132<br>2.131<br>2.192 | **Bi**<br>1.979<br>1.987<br>1.983<br>2.035 | **Po**<br>1.845<br>1.858<br>1.845<br>1.902 | **At**<br>2.652<br>2.646<br>2.647<br>2.741 | **Rn**<br>3.637<br>3.284<br>3.385<br>3.454 |

## 3. Adsorption energies at the single vacancy at graphene sheet

**Table S2.** Calculated adsorption energies of atoms in PTE on SV−graphene (in eV).

| | | | | | | | | | | | | | | | | | |
|---|---|---|---|---|---|---|---|---|---|---|---|---|---|---|---|---|---|
| **H**<br>−3.29<br>−3.37<br>−3.33<br>**−3.41** | | | | | **element**<br>PBE<br>PBE+D2<br>PBE+D3<br>vdW−DF2 | | | | | | | | | | | | **He**<br>−0.01<br>−0.02<br>**−0.03**<br>−0.02 |
| **Li**<br>−2.65<br>**−2.94**<br>−2.72<br>−2.59 | **Be**<br>−6.17<br>**−6.30**<br>−6.23<br>−6.29 | | | | | | | | | | | **B**<br>−12.91<br>**−13.04**<br>−12.96<br>−11.76 | **C**<br>−15.60<br>**−15.70**<br>−15.64<br>−15.60 | **N**<br>−12.02<br>**−12.11**<br>−12.07<br>−9.51 | **O**<br>−7.77<br>**−7.85**<br>−7.83<br>−5.41 | **F**<br>−4.55<br>**−4.67**<br>−4.62<br>−2.43 | **Ne**<br>nb<br>−0.04<br>−0.03<br>−0.04 |
| **Na**<br>−1.88<br>**−2.20**<br>−2.00<br>−1.74 | **Mg**<br>−1.70<br>**−1.98**<br>−1.80<br>−1.40 | | | | | | | | | | | **Al**<br>−5.15<br>**−5.37**<br>−5.23<br>−4.75 | **Si**<br>−8.21<br>**−8.38**<br>−8.28<br>−7.62 | **P**<br>−8.31<br>**−8.47**<br>−8.41<br>−7.70 | **S**<br>−7.13<br>**−7.27**<br>−7.25<br>−6.08 | **Cl**<br>−2.85<br>**−3.03**<br>−3.01<br>−2.51 | **Ar**<br>−0.00<br>−0.08<br>**−0.09**<br>−0.08 |
| **K**<br>−2.09<br>**−2.36**<br>−2.22<br>−1.92 | **Ca**<br>−3.18<br>**−3.50**<br>−3.27<br>−2.78 | **Sc**<br>−6.48<br>**−6.77**<br>−6.59<br>−5.89 | **Ti**<br>−8.03<br>**−8.30**<br>−8.14<br>−7.22 | **V**<br>−7.27<br>**−7.54**<br>−7.38<br>−6.87 | **Cr**<br>−6.16<br>**−6.42**<br>−6.28<br>−5.95 | **Mn**<br>−6.18<br>**−6.43**<br>−6.31<br>−5.56 | **Fe**<br>−7.10<br>**−7.35**<br>−7.23<br>−6.32 | **Co**<br>−7.68<br>**−7.92**<br>−7.79<br>−6.64 | **Ni**<br>−6.64<br>**−6.89**<br>−6.78<br>−5.72 | **Cu**<br>−3.61<br>**−3.87**<br>−3.75<br>−2.89 | **Zn**<br>−1.05<br>**−1.32**<br>−1.18<br>−0.30 | **Ga**<br>−3.96<br>**−4.24**<br>−4.07<br>−3.08 | **Ge**<br>−6.30<br>**−6.54**<br>−6.39<br>−5.59 | **As**<br>−6.62<br>**−6.84**<br>−6.74<br>−6.04 | **Se**<br>−5.60<br>**−5.79**<br>−5.74<br>−4.61 | **Br**<br>−2.20<br>**−2.43**<br>−2.39<br>−1.89 | **Kr**<br>−0.01<br>−0.12<br>−0.12<br>**−0.13** |
| **Rb**<br>−2.06<br>**−2.41**<br>−2.23<br>−1.88 | **Sr**<br>−2.83<br>**−3.23**<br>−2.93<br>−2.39 | **Y**<br>−6.21<br>**−6.60**<br>−6.32<br>−5.57 | **Zr**<br>−8.34<br>**−8.71**<br>−8.45<br>−7.51 | **Nb**<br>−8.15<br>**−8.52**<br>−8.26<br>−7.74 | **Mo**<br>−7.29<br>**−7.65**<br>−7.43<br>−6.93 | **Tc**<br>−8.26<br>**−8.61**<br>−8.47<br>−7.31 | **Ru**<br>−8.81<br>**−9.16**<br>−8.99<br>−7.67 | **Rh**<br>−8.34<br>**−8.69**<br>−8.49<br>−7.05 | **Pd**<br>−5.27<br>**−5.62**<br>−5.44<br>−4.30 | **Ag**<br>−1.72<br>**−2.11**<br>−1.89<br>−1.28 | **Cd**<br>−0.10<br>**−0.47**<br>−0.34<br>−0.19 | **In**<br>−2.12<br>**−2.58**<br>−2.28<br>−1.94 | **Sn**<br>−4.28<br>**−4.64**<br>−4.39<br>−3.96 | **Sb**<br>−4.98<br>**−5.29**<br>−5.01<br>−4.72 | **Te**<br>−4.24<br>**−4.51**<br>−4.42<br>−3.41 | **I**<br>−1.67<br>**−1.97**<br>−1.89<br>−1.44 | **Xe**<br>−0.03<br>−0.15<br>**−0.16**<br>−0.11 |
| **Cs**<br>−2.20<br>**−3.33**<br>−2.37<br>−2.01 | **Ba**<br>−3.38<br>**−4.39**<br>−3.49<br>−2.90 | **La**<br>−6.70<br>**−7.62**<br>−6.81<br>−5.98 | **Hf**<br>−8.60<br>**−9.17**<br>−8.71<br>−7.64 | **Ta**<br>−8.99<br>**−9.50**<br>−9.13<br>−7.92 | **W**<br>−8.52<br>**−9.03**<br>−8.72<br>−8.00 | **Re**<br>−8.13<br>**−8.63**<br>−8.37<br>−7.31 | **Os**<br>−9.29<br>**−9.79**<br>−9.53<br>−8.03 | **Ir**<br>−9.28<br>**−9.77**<br>−9.45<br>−7.67 | **Pt**<br>−7.08<br>**−7.57**<br>−7.34<br>−6.02 | **Au**<br>−2.40<br>**−2.93**<br>−2.60<br>−1.80 | **Hg**<br>−0.03<br>**−0.40**<br>−0.29<br>−0.19 | **Tl**<br>−2.12<br>**−2.50**<br>−2.38<br>−1.96 | **Pb**<br>−3.31<br>**−3.69**<br>−3.46<br>−3.10 | **Bi**<br>−4.06<br>**−4.46**<br>−4.19<br>−3.83 | **Po**<br>−3.38<br>**−3.68**<br>−3.58<br>−2.56 | **At**<br>−1.41<br>**−1.77**<br>−1.67<br>−1.22 | **Rn**<br>−0.00<br>**−0.24**<br>−0.18<br>−0.15 |

## 4. Magnetic properties of A@SV-graphene systems

The magnetic properties of M@SV-graphene systems are given in Table S3. The applied computational schemes agree well considering predicted magnetic moments. However, this property is rather sensitive to the computational treatment and it has been shown for the case of Fe that the use of DFT+$U$ gives rise to a magnetic ground state which is not the case with PBE and, as we show, vdW-DF2.[1] For a detailed analysis of bonding in the case of Fe, which can be further translated to the cases of Ru and Os, the reader is referred to ref. ([1]). However, in spite the problems in prediction of magnetism in M@SV-graphene systems associated with PBE, it can be concluded that the adsorption of foreign atoms into SV site alters magnetic properties of SV-graphene to a great extent. Our results suggest that large fraction of considered atoms results with non-magnetic ground states, while some elements (alkaline and earth alkaline metals, elements in the middle of the d-block, metals in groups 11 and 12) give magnetic systems with appreciable magnetic moments.

**Table S3.** Ground state magnetizations of A@v−G systems (in Bohr magnetons, $\mu_B$).

| | | | | | | | | | | | | | | | | | |
|---|---|---|---|---|---|---|---|---|---|---|---|---|---|---|---|---|---|
| **H**<br>2.74<br>2.74<br>2.74<br>2.48 | | | | | **element**<br>PBE<br>PBE+D2<br>PBE+D3<br>vdW−DF2 | | | | | | | | | | | | **He**<br>1.36<br>1.36<br>1.36<br>1.38 |
| **Li**<br>0.82<br>0.82<br>0.82<br>2.20 | **Be**<br>0.00<br>0.00<br>0.00<br>0.00 | | | | | | | | | | | **B**<br>0.00<br>0.00<br>0.00<br>0.00 | **C**<br>0.00<br>0.00<br>0.00<br>0.00 | **N**<br>0.00<br>0.00<br>0.00<br>0.00 | **O**<br>0.00<br>0.00<br>0.00<br>0.00 | **F**<br>0.00<br>0.00<br>0.00<br>0.00 | **Ne**<br>1.36<br>1.37<br>1.36<br>1.38 |
| **Na**<br>0.98<br>0.98<br>0.98<br>0.98 | **Mg**<br>1.94<br>1.94<br>1.94<br>1.89 | | | | | | | | | | | **Al**<br>0.00<br>0.00<br>0.00<br>0.00 | **Si**<br>0.00<br>0.00<br>0.00<br>0.00 | **P**<br>0.00<br>0.00<br>0.00<br>0.00 | **S**<br>0.00<br>0.00<br>0.00<br>0.00 | **Cl**<br>0.01<br>0.00<br>0.00<br>0.39 | **Ar**<br>1.36<br>1.37<br>1.37<br>1.45 |
| **K**<br>1.00<br>1.00<br>1.00<br>0.99 | **Ca**<br>1.89<br>1.9<br>1.89<br>1.82 | **Sc**<br>0.00<br>0.00<br>0.00<br>0.00 | **Ti**<br>0.00<br>0.00<br>0.00<br>0.00 | **V**<br>1.01<br>1.01<br>1.01<br>1.01 | **Cr**<br>2.01<br>2.02<br>2.01<br>2.05 | **Mn**<br>2.84<br>2.83<br>2.84<br>2.76 | **Fe**<br>0.00<br>0.00<br>0.00<br>0.00 | **Co**<br>0.28<br>0.28<br>0.28<br>0.00 | **Ni**<br>0.00<br>0.00<br>0.00<br>0.00 | **Cu**<br>0.87<br>0.77<br>0.93<br>0.94 | **Zn**<br>1.00<br>1.18<br>1.12<br>1.78 | **Ga**<br>0.00<br>0.00<br>0.00<br>0.00 | **Ge**<br>0.00<br>0.00<br>0.00<br>0.00 | **As**<br>0.00<br>0.00<br>0.00<br>0.00 | **Se**<br>0.00<br>0.00<br>0.00<br>0.00 | **Br**<br>0.00<br>0.00<br>0.00<br>0.00 | **Kr**<br>1.36<br>1.37<br>1.37<br>1.38 |
| **Rb**<br>1.00<br>1.00<br>1.00<br>1.00 | **Sr**<br>1.96<br>1.96<br>1.96<br>1.93 | **Y**<br>0.00<br>0.00<br>0.00<br>0.00 | **Zr**<br>0.00<br>0.00<br>0.00<br>0.00 | **Nb**<br>0.97<br>0.97<br>0.97<br>0.97 | **Mo**<br>1.98<br>1.98<br>1.98<br>1.99 | **Tc**<br>0.94<br>0.94<br>0.93<br>0.95 | **Ru**<br>0.00<br>0.00<br>0.00<br>0.00 | **Rh**<br>0.00<br>0.00<br>0.00<br>0.00 | **Pd**<br>0.00<br>0.00<br>0.00<br>0.00 | **Ag**<br>1.02<br>1.00<br>1.00<br>0.95 | **Cd**<br>1.69<br>1.72<br>1.50<br>1.31 | **In**<br>2.33<br>2.34<br>2.30<br>2.22 | **Sn**<br>0.00<br>0.00<br>0.00<br>0.00 | **Sb**<br>0.00<br>0.00<br>0.00<br>0.00 | **Te**<br>0.00<br>0.00<br>0.00<br>0.00 | **I**<br>0.02<br>0.07<br>0.15<br>0.60 | **Xe**<br>1.37<br>1.42<br>1.39<br>1.41 |
| **Cs**<br>1.00<br>1.00<br>1.00<br>1.00 | **Ba**<br>1.97<br>1.98<br>1.96<br>1.94 | **La**<br>0.00<br>0.00<br>0.00<br>0.00 | **Hf**<br>0.00<br>0.00<br>0.00<br>0.00 | **Ta**<br>0.75<br>0.77<br>0.76<br>0.63 | **W**<br>1.97<br>1.97<br>1.97<br>1.99 | **Re**<br>0.94<br>0.93<br>0.94<br>1.02 | **Os**<br>0.00<br>0.00<br>0.00<br>0.00 | **Ir**<br>0.71<br>0.71<br>0.71<br>0.00 | **Pt**<br>0.00<br>0.00<br>0.00<br>0.00 | **Au**<br>1.00<br>1.00<br>1.00<br>0.59 | **Hg**<br>1.34<br>1.40<br>1.38<br>1.35 | **Tl**<br>0.81<br>0.80<br>0.82<br>0.77 | **Pb**<br>0.00<br>0.00<br>0.00<br>0.00 | **Bi**<br>0.00<br>0.00<br>0.00<br>0.00 | **Po**<br>0.00<br>0.00<br>0.00<br>0.00 | **At**<br>0.37<br>0.36<br>0.41<br>0.71 | **Rn**<br>1.37<br>1.41<br>1.40<br>1.46 |

An interesting situation is seen in the case of adsorption of noble gases. In all cases we observe magnetism which is in magnitude very close to that of the pure SV-graphene substrate. For this reason we checked spin polarization in these systems (see a representative picture in Fig. S3) and confirmed that the magnetism is due to the vacancy itself. Due to very weak interactions of noble gases with the vacancy it is almost unaffected compared to the pure SV-graphene.

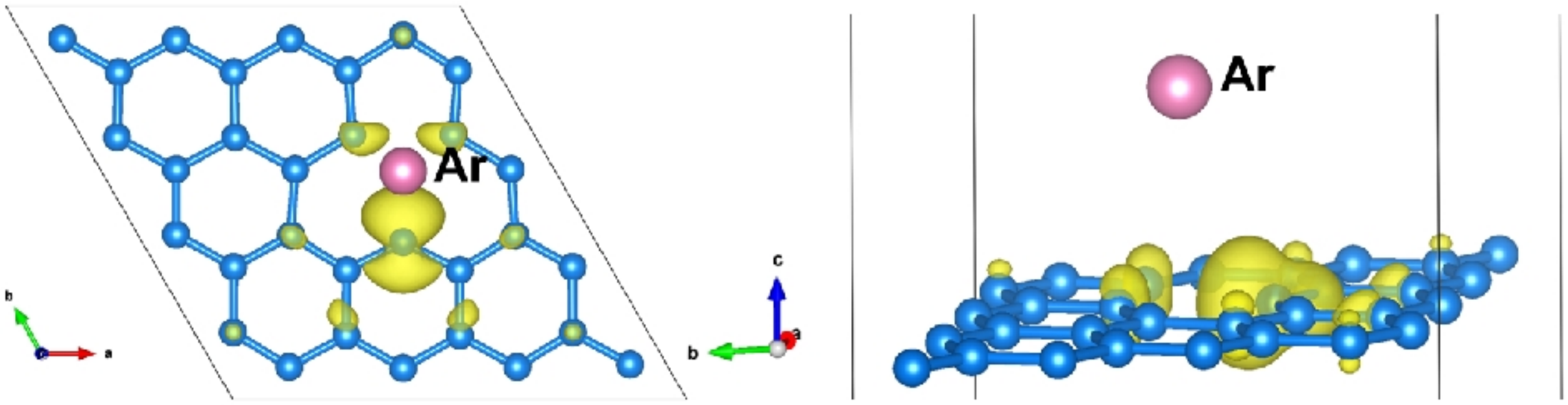

**Figure S3.** Distribution of spin polarization ($\rho_{\text{spin up}} - \rho_{\text{spin down}}$) for the ground states of Ar at graphene single vacancy calculated at the PBE level (the isovalues are ±0.006 e Å$^{-3}$).

## 5. Correlation of adsorption energies

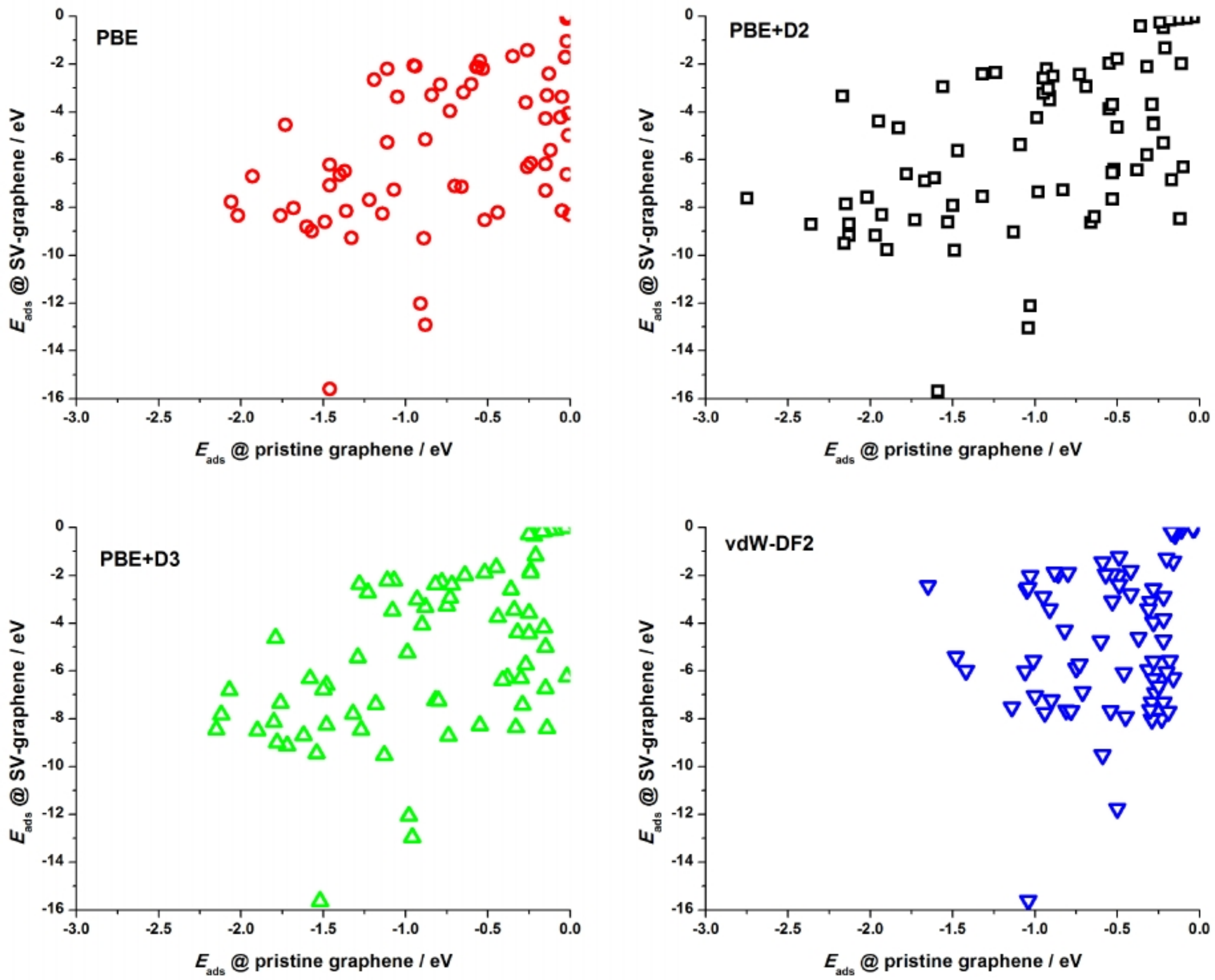

**Figure S4**. Correlation between the calculated adsorption energies of investigated elements on pristine graphene and SV-graphene, for the four methods applied.

## 6. Dissolution potentials of metals from the SV site of graphene

We have calculated electrode potentials for the reaction

$M^{z+}$ + SV-graphene + $ze^-$ → M@SV-graphene (S1)

denoted hereafter as $E^{o}_{M^{z+}/M@SV\text{-graphene}}$. The potentials were calculated on the basis of the standard electrode potential ( $E^{o}_{M^{z+}/M}$ ) for the reaction:

$M^{z+} + ze^- \rightarrow M_{(s)}$ (S2)

assuming a galvanic cell where anode is made of pure metal M and the cathode is made of SV-graphene, where the half-reaction is given by Eq. (S1). Activity of $M^{z+}$ ions is assumed to be one. Assuming that there are no volume changes during the reaction and that the entropy contribution can be disregarded, $E^{o}_{M^{z+}/M@SV\text{-graphene}}$ is given by:

$$E^{o}_{M^{z+}/M@SV\text{-graphene}} = E^{o}_{M^{z+}/M} - \frac{E_{ads} + E_{coh}}{zF} \quad (S3)$$

In the above equation $F$ if Faraday constant while $E_{ads}$ and $E_{coh}$ should be expressed in [J mol$^{-1}$] The calculated potentials (Table S4) are referred to the Standard Hydrogen Electrode (SHE) at 25°C.

**Table S4**. Calculated values of $E^{o}_{M^{z+}/M@SV\text{-graphene}}$ for the four methods applied. Where the half-reaction (S2) can take place with different $z$ we always took the most negative $E^{o}_{M^{z+}/M}$. Elements are assembled as they appear in the groups of the PTE, with the increasing atomic number.

| | | | $E^{o}_{M^{z+}/M@SV\text{-graphene}}$ / V | | | |
|---|---|---|---|---|---|---|
| **metal** | $E^{o}_{M^{z+}/M}$ / V | $z$ | **PBE** | **PBE+D2** | **PBE+D3** | **vdW-DF2** |
| Li | −3.0401 | 1 | −2.02 | −1.73 | −1.95 | −2.08 |
| Na | −2.71 | 1 | −1.94 | −1.62 | −1.82 | −1.82 |
| K | −2.931 | 1 | −1.78 | −1.51 | −1.65 | −1.95 |
| Rb | −2.98 | 1 | −1.77 | −1.42 | −1.60 | −1.95 |
| Cs | −3.206 | 1 | −1.81 | −0.68 | −1.64 | −2.00 |
| Be | −1.847 | 2 | −0.42 | −0.36 | −0.39 | −0.36 |
| Mg | −2.372 | 2 | −2.28 | −2.14 | −2.23 | −2.43 |
| Ca | −2.868 | 2 | −2.20 | −2.04 | −2.15 | −2.40 |
| Sr | −2.889 | 2 | −2.33 | −2.13 | −2.28 | −2.55 |
| Ba | −2.912 | 2 | −2.17 | −1.67 | −2.12 | −2.41 |
| Sc | −2.077 | 3 | −1.22 | −1.12 | −1.18 | −1.41 |
| Y | −2.372 | 3 | −1.76 | −1.63 | −1.72 | −1.97 |
| La | −2.379 | 3 | −1.64 | −1.33 | −1.60 | −1.88 |
| Ti | −1.63 | 2 | −0.04 | 0.10 | 0.02 | −0.45 |
| Zr | −1.45 | 4 | −0.93 | −0.84 | −0.90 | −1.14 |
| Hf | −1.55 | 4 | −1.01 | −0.87 | −0.98 | −1.25 |
| V | −1.13 | 2 | −0.15 | −0.01 | −0.10 | −0.35 |
| Nb | −1.099 | 3 | −0.91 | −0.78 | −0.87 | −1.04 |
| Ta | −0.60 | 3 | −0.30 | −0.13 | −0.26 | −0.66 |
| Cr | −0.74 | 3 | −0.05 | 0.03 | −0.01 | −0.12 |
| Mo | no data | | | | | |
| W | no data | | | | | |
| Mn | −1.185 | 2 | 0.45 | 0.57 | 0.51 | 0.14 |
| Tc | no data | | | | | |
| Re | 0.30 | 3 | 0.33 | 0.50 | 0.41 | 0.06 |
| Fe | −0.44 | 2 | 0.97 | 1.10 | 1.04 | 0.58 |
| Ru | 0.60 | 3 | 1.29 | 1.41 | 1.35 | 0.91 |
| Os | no data | | | | | |
| Co | −0.28 | 2 | 1.37 | 1.49 | 1.42 | 0.85 |
| Rh | 0.76 | 3 | 1.62 | 1.74 | 1.67 | 1.19 |
| Ir | no data | | | | | |
| Ni | −0.25 | 2 | 0.85 | 0.98 | 0.92 | 0.39 |
| Pd | 0.915 | 2 | 1.61 | 1.78 | 1.69 | 1.12 |
| Pt | 1.188 | 2 | 1.81 | 2.05 | 1.94 | 1.28 |
| Cu | 0.337 | 2 | 0.40 | 0.53 | 0.47 | 0.04 |
| Ag | 0.7996 | 1 | −0.43 | −0.04 | −0.26 | −0.87 |
| Au | 1.52 | 3 | 1.26 | 1.44 | 1.33 | 1.06 |
| Zn | −0.7618 | 2 | −0.91 | −0.78 | −0.85 | −1.29 |
| Cd | −0.40 | 2 | −0.93 | −0.75 | −0.81 | −0.89 |
| Hg | 0.85 | 2 | 0.53 | 0.72 | 0.66 | 0.61 |
| Al | −1.662 | 3 | −1.08 | −1.00 | −1.05 | −1.21 |
| Ga | −0.53 | 3 | −0.15 | −0.05 | −0.11 | −0.44 |
| In | −0.34 | 3 | −0.47 | −0.32 | −0.42 | −0.53 |
| Tl | −0.34 | 1 | −0.10 | 0.28 | 0.16 | −0.26 |
| Sn | −0.13 | 2 | 0.44 | 0.62 | 0.50 | 0.28 |
| Pb | −0.126 | 2 | 0.51 | 0.70 | 0.59 | 0.41 |
| As | no data | | | | | |
| Sb | no data | | | | | |
| Bi | 0.308 | 3 | 0.93 | 1.07 | 0.98 | 0.86 |
| Te | −0.9 | 2 | 0.13 | 0.26 | 0.22 | −0.29 |
| Po | no data | | | | | |

## Supplementary references